\documentclass{emulateapj}
\slugcomment{}

\begin{document}

\shortauthors{Lin et al.}
\shorttitle{LMXB and IMXB Evolution}
\title{LMXB and IMXB Evolution: I. The Binary Radio Pulsar PSR J1614-2230}

\author{Jinrong Lin\altaffilmark{1}, S. Rappaport\altaffilmark{1}, Ph. Podsiadlowski\altaffilmark{2},
L. Nelson\altaffilmark{3}, B. Paxton\altaffilmark{4}, \& P. Todorov\altaffilmark{5}}

\altaffiltext{1}{37-624B, M.I.T. Department of Physics and Kavli
Institute for Astrophysics and Space Research, 70 Vassar St.,
Cambridge, MA, 02139; jinrongl@mit.edu; sar@mit.edu}

\altaffiltext{2}{Department of Astrophysics, Oxford University,
Oxford OX1 3RH, UK; podsi@astro.ox.ac.uk}

\altaffiltext{3}{Department of Physics, Bishops University,
Sherbrooke, QC J1M 1Z7, Canada; lnelson@ubishops.ca}

\altaffiltext{4}{KITP, Kohn Hall, University of California at
Santa Barbara, CA 93106-4030; paxton@kitp.ucsb.edu}

\altaffiltext{5}{Laboratoire de l'Univers et ses Th\'eories,
Observatoire de Paris, 5 place Jules Janssens, F-92190 Meudon
Cedex; petar.todorov@obspm.fr}

\begin{abstract}
We have computed an extensive grid of binary evolution tracks to
represent low- and intermediate mass X-ray binaries (LMXBs and
IMXBs). The grid includes 42,000 models which covers 60 initial
donor masses over the range of $1-4\,M_\odot$ and, for each of
these, 700 initial orbital periods over the range of $10-250$
hours. These results can be applied to understanding LMXBs and
IMXBs: those that evolve analogously to CVs; that form
ultracompact binaries with $P_{\rm orb}$ in the range of $6-50$
minutes; and that lead to wide orbits with giant donors. We also
investigate the relic binary recycled radio pulsars into which
these systems evolve. To evolve the donor stars in this study, we
utilized a newly developed stellar evolution code called ``{\tt
MESA}'' that was designed, among other things, to be able to
handle very low-mass and degenerate donors. This first application
of the results is aimed at an understanding of the newly
discovered pulsar PSR J1614-2230 which has a $1.97\,M_\odot$
neutron star, $P_{\rm orb} = 8.7$ days, and a companion star of
$0.5\,M_\odot$. We show that (i) this system is a cousin to the
LMXB Cyg X-2; (ii) for neutron stars of canonical birth mass
$1.4\,M_\odot$, the initial donor stars which produce the closest
relatives to PSR J1614-2230 have a mass between
$3.4-3.8\,M_\odot$; (iii) neutron stars as massive as $1.97 \,
M_\odot$ are not easy to produce in spite of the initially high
mass of the donor star, unless they were already born as
relatively massive neutron stars; (iv) to successfully produce a
system like PSR J1614-2230 requires a minimum initial neutron star
mass of at least $1.6\pm 0.1 \,M_\odot$, as well as initial donor
masses and $P_{\rm orb}$ of $\sim$$4.25 \pm 0.10 \,M_\odot$ and
$\sim$$49 \pm 2$ hrs, respectively; and (v) the current companion
star is largely composed of CO, but should have a surface H
abundance of $\sim$$10-15\%$.
\end{abstract}

\keywords{stars: binaries: general --- stars: evolution ---
 stars: pulsars: individual (PSR J1614-2230) --- accretion,
accretion disks --- X-rays: binaries}

\section{Introduction}

There has been an-ongoing effort to understand the evolution of
low-mass X-ray binaries (LMXBs) since their basic nature was
understood back in the late 1960's and early 1970's \citep[see,
e.g.,][]{faul71,rapp83,webb83,joss84,nels85,pyly88,pyly89,bhat91,iben95,pods02,pfah03,nels04,belc08}.
There are two rather distinct parts of the evolution to consider:
(i) the formation of a neutron star in orbit with a low-mass donor
star, and (ii) the subsequent portion of the evolution when mass
is transferred from the companion donor star to the neutron star.

The difficulty with the first part of the evolution involves the
conceptual problem of keeping the binary system bound while the
massive progenitor of the neutron star (NS) explodes in a
supernova event. A part of this problem was addressed by invoking
a common envelope phase during which the lower-mass secondary
ejects the envelope of the massive NS progenitor
\citep{pacz76,meye79,webb79,bhat91}. However, even this solution
is not so straightforward in that there may be insufficient
gravitational energy release between the inspiraling low-mass star
and the core of the NS progenitor to successfully eject the
envelope \citep[see, e.g.,][]{dewi00,pfah03}. This difficulty
could be overcome by invoking secondaries that are of intermediate
mass, e.g., $2-4\, M_\odot$ to increase the gravitational energy
release \citep[see, e.g.,][]{pfah03}. In turn, there has been a
persistent conceptual misunderstanding that the transfer of mass
from a star of $2-4\,M_\odot$ onto a $1.4 \,M_\odot$ NS was
dynamically unstable. That was shown to be untrue
(\citealp{pyly88,pyly89,taur99,pods00,king01}; PRP02; see also
\citealp{davi98,king99}).

Starting in the 1970's and continuing until the present, there
have been numerous evolution studies of a limited number of LMXBs
and IMXBs exploring the various paths which lead to very different
intermediate and end-stage products. One of the more systematic
was the study by \citet{pods02} and \citet{pfah03} which involved
about 150 LMXB and IMXB systems. Given the two-dimensional
parameter space of initial $P_{\rm orb,i}$ and $M_{\rm 2,i}$
(where $P_{\rm orb,i}$ and $M_{\rm 2,i}$ are the initial orbital
period and mass of the donor star), the types of evolutionary
paths for LMXBs or IMXBs is really quite large and varied.
Possibilities include evolution to a minimum orbital period of
$\sim$70 min with very low-mass H-rich donor stars; evolution to
an ultracompact state with $P_{\rm orb}$ in the range of $6-50$
minutes and He-rich donors; and evolution to wide orbits of
days-to-months with low-mass giant donor stars.

In order to better understand the many possible evolution paths of
LMXBs and IMXBs in a more systematic way, we have calculated an
extensive grid of binary models encompassing 42,000 initial
combinations of $P_{\rm orb,i}$ and $M_{\rm 2,i}$. This is two
orders of magnitude larger than the study we conducted in 2002. We
took advantage of a newly developed stellar evolution code whose
equations of state allow for the evolution of very low mass stars
with cold dense interiors, and a cluster of computers which speeds
up the overall calculation beyond what was readily available a
decade ago.

With the recent discovery of PSR J1614-2230, with the most massive
neutron star known ($1.97\pm 0.04\,M_\odot$), a relatively close
8.7-day orbit, and a fairly massive white dwarf companion
($0.5\,M_\odot$), we decided to first apply our evolution
calculations to understanding the origins of this system. The
specific goals are to understand whether a natal NS with canonical
mass of $\sim$$1.4\,M_\odot$ can accrete sufficient material to
grow to nearly $2\,M_\odot$, to see whether the donor star mass is
consistent with the observed $P_{\rm orb}$, and to investigate how
this system is related to possible progenitors in the guise of the
LMXB -- Cyg X-2. In all, we found $\sim$500 of our evolution
tracks which produce systems that are at least generically related
to PSR J1614-2230.

In this paper we introduce our set of binary evolution
calculations which cover an extensive grid of initial orbital
periods and donor masses (\S \ref{sec:overview}). In \S
\ref{sec:PSR1614} we show how our binary evolution calculations
apply directly to PSR J1614-2230. In \S \ref{sec:Discuss} we
discuss a number of the general lessons we have gleaned from this
study.

\section{Overview of LMXB and IMXB Binary Evolution}
\label{sec:overview}

\subsection{Binary Evolution Calculations }

In this work we start with binary systems with an unevolved
companion star in a circular orbit with an already formed neutron
star of mass 1.4 $M_\odot$. The prior evolution, leading to the NS
and low- or intermediate-mass donor star does not concern us in
this study. However, this prior phase of evolution involves a much
more massive progenitor of the NS (perhaps $8-15\,M_\odot$), as
well as a common envelope phase that unveils the He/CO core of the
progenitor which, in turn, evolves to collapse and the formation
of a neutron star (\citealp[see, e.g.,][]{bhat91,kalo98,taur00},
\citealp[and references therein]{will02}; \citealp{pfah03}).

The subsequent evolution of the incipient LMXB or IMXB is computed
with a combination of a newly developed Henyey code called ``{\tt
MESA}'' (Modules for Experiments in Stellar Astrophysics;
\citealp{paxt10}) and a binary driver code. The advantages of {\tt
MESA} for our calculations are that (i) it is specifically
designed to include equations of state that are able to handle the
low-mass and high degree of degeneracy that the donor stars
achieve (see \citealp{paxt10} for details), and (ii) it is robust
when used in a binary evolution code that allows for `hands-off'
evolution through all the various stages of the donor star and
phases of mass transfer. We found that none of the evolution
models failed to run to their expected completion.

\begin{figure*}[t]\epsscale{0.8}
\plotone{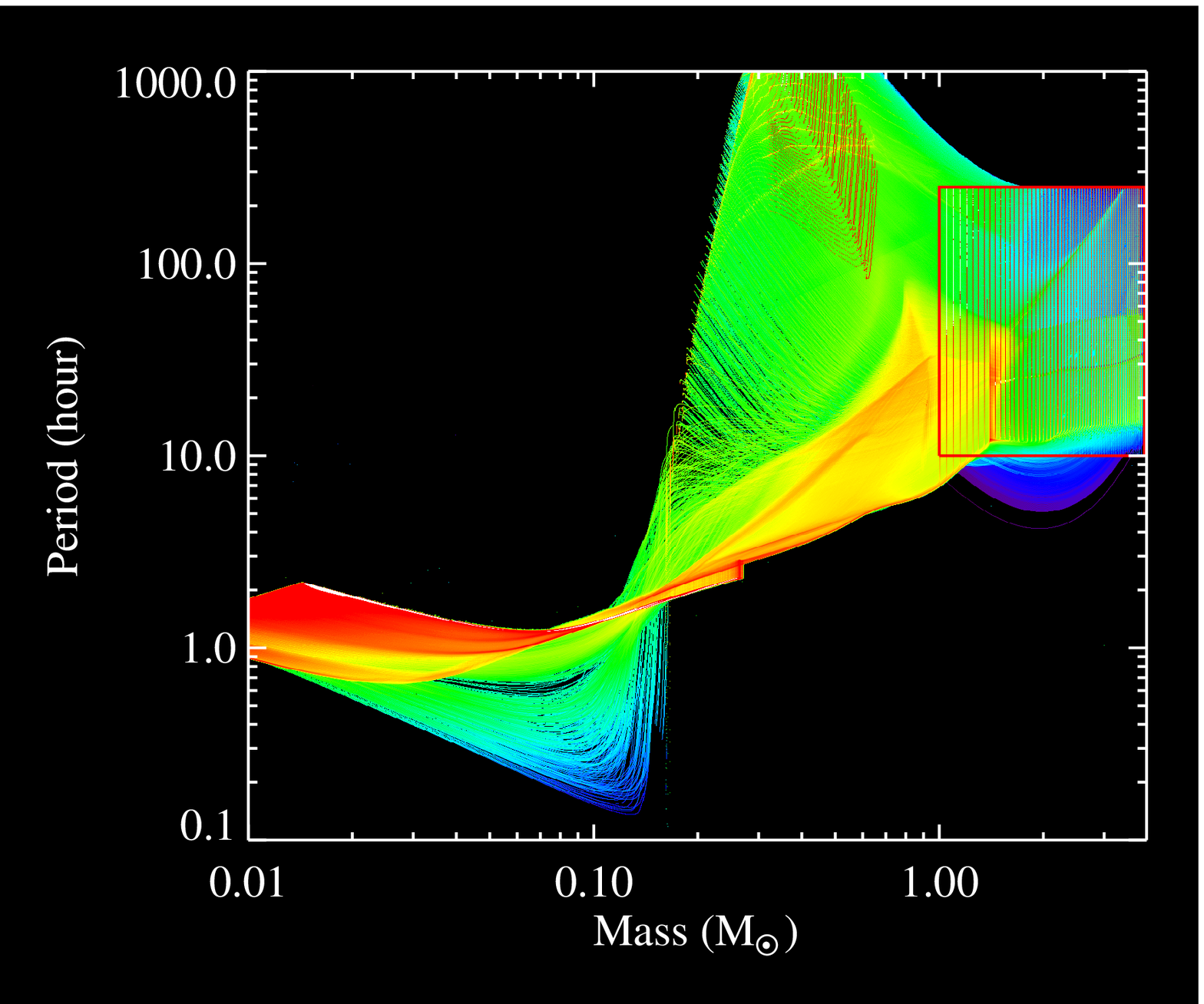} \caption{Superposed set of 38,000 binary
evolution tracks in the $P_{\rm orb}-M_2$ plane. Each time a track
crosses one of the $1300 \times 1200$ grid points in the image,
the evolution time step is stored in that pixel. The color scaling
is related to the logarithm of the total system time spent in each
pixel. The accumulated evolution times (per pixel) range from
$10^3$ yr for purple to $10^{10}$ yr for red. The initial grid of
models is visible in the upper right portion of the plot, outlined
by a red box.} \label{fig:43000}

\vspace{0.5cm}

\plotone{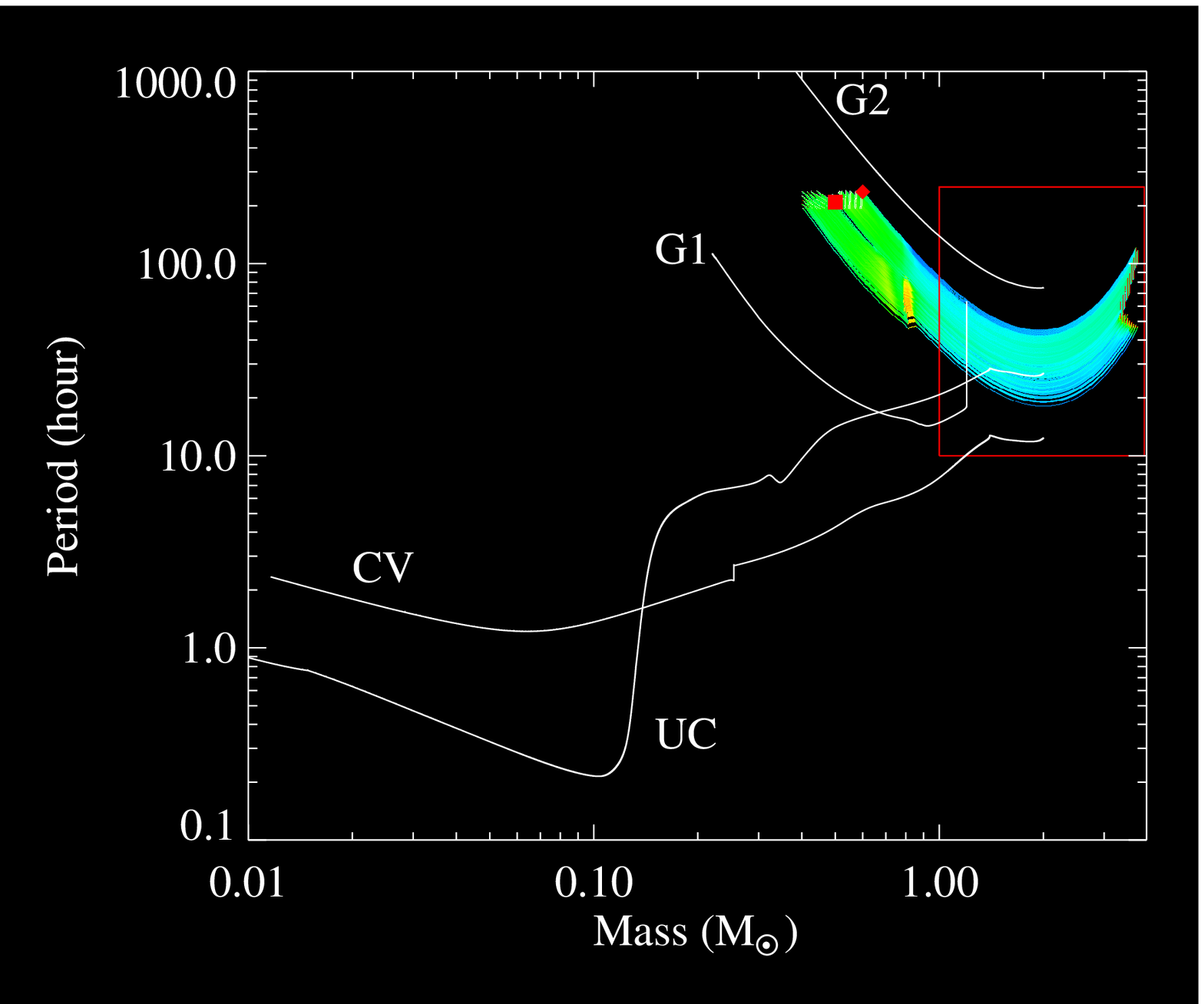} \caption{Evolution tracks of all 515 systems
that terminated their evolution with $P_{\rm orb} = 9\pm 1$ day,
and $M_2 = 0.5 \pm 0.1 M_\odot$. The colors span cumulative
evolution times (per pixel) of between $10^4$ yr (light blue) to
$10^5$ yr (green). The small yellow region represents H-burning
for up to $3 \times 10^6$ yr (per pixel). The location of PSR
J1614-2230 is marked with a red square; that of Cyg X-2 with a red
diamond.  The four white tracks represent other typical evolutionary
paths for L/IMXBs (see text for details,)} \label{fig:Ransom_like}
\end{figure*}

The binary driver code evolves the binary orbit, including the
effects of mass transfer and angular momentum losses due to
gravitational radiation, magnetic braking, and possible mass
ejection from the system. It keeps track of where the Roche lobe
is in respect to the atmosphere of the donor star being evolved by
{\tt MESA}, and decides how much mass to remove from the donor
star during each evolution time step. It also manages the various
disparate timescales involved in the donor star as well as in the
evolution of the binary orbit (details of the driver code will be
presented in a future work, but its operation is very similar to
that described in \citealp{madh08}). For the specific models
presented here, we assumed that for sub-Eddington mass transfer
rates, 90\% of the mass transferred is retained by the neutron
star, but that for higher transfer rates the accretion onto the NS
was limited to that set by the Eddington limit. Matter ejected
from the system was assumed to carry away the specific angular
momentum of the neutron star.

No tidal evolution (between the donor star's rotation and the
orbital angular momentum) was incorporated in the code.  We regard
this as typically a $\lesssim 10\%$ effect on the evolution, both
before mass transfer commences and afterward. Magnetic braking is
included (according to eq.\,[36] in \citealp{rapp83}) for all
stars except those that are completely convective or have a mass
$>1.4\,M_\odot$. As an approximation, we also assume that magnetic
braking operates even in those cases where the donor star
underfills its Roche lobe and continuous synchronization between
the donor star and the orbit is not guaranteed.  We have verified
by numerical experiment that this approximation makes little
difference in the final distribution of evolution tracks shown in
Fig.\,\ref{fig:43000}. In this work we do not consider either
thermal-ionization disk instabilities or the X-ray irradiation of
the donor star during the course of the evolution calculations.
Their effects can be approximated via the application of
after-the-fact algorithms -- which we discuss in a future work.
However, for the purposes of the calculations emphasized in this
paper, neither effect is very important.

We computed binary evolution tracks for $42,000$ LMXBs and IMXBs
over a grid of 60 initial donor masses uniformly distributed over
the range of $1-4\,M_\odot$ and 700 initial orbital periods
distributed over the range of $10-250$ hours in equal logarithmic
steps. With our dense coverage of the initial binary parameter
space, most of the basic types of LMXB and IMXB evolutionary
tracks are explored.

The ``initial orbital period'' in these runs is defined as $P_{\rm
orb}$ following the birth of the neutron star. The orbits are
assumed to be circular, and the donor stars start on the ZAMS in
our evolution calculations. The binary evolution is considered
`complete' when either (i) 10 Gyr have elapsed, or (ii) the mass
transfer becomes dynamically unstable (this later condition occurs
for only $\sim$2.4\% of our systems, typically the ones with the
largest initial donor masses and orbital periods). After the
envelope of the donor star has been transferred to the neutron
star (or ejected from the system), {\tt MESA} continues to evolve
the relic He or He/CO core of the donor until a total elapsed time
of 10 Gyr has passed.

\subsection{Results}

A graphic presentation of our 38,000 binary evolution tracks,
representing LMXBs and IMXBs is shown in Fig.\,\ref{fig:43000} in
the $P_{\rm orb}-M_2$ plane. We divide this plane up into $1300
\times 1200$ discrete pixels. As the evolution tracks pass through
this plane, we record, in a cumulative fashion, the evolution time
that is spent in crossing a given pixel. When all the tracks have
been co-added in this fashion, we display the logarithm of the
total accumulated time in a pixel with color shading. The
cumulative dwell times per pixel range from $10^{10}$ yr for the
red regions to $10^3$ yr for the purple. The red box in the upper
right of the diagram outlines the starting grid of initial models.

\begin{figure*}\epsscale{1.15}
\plotone{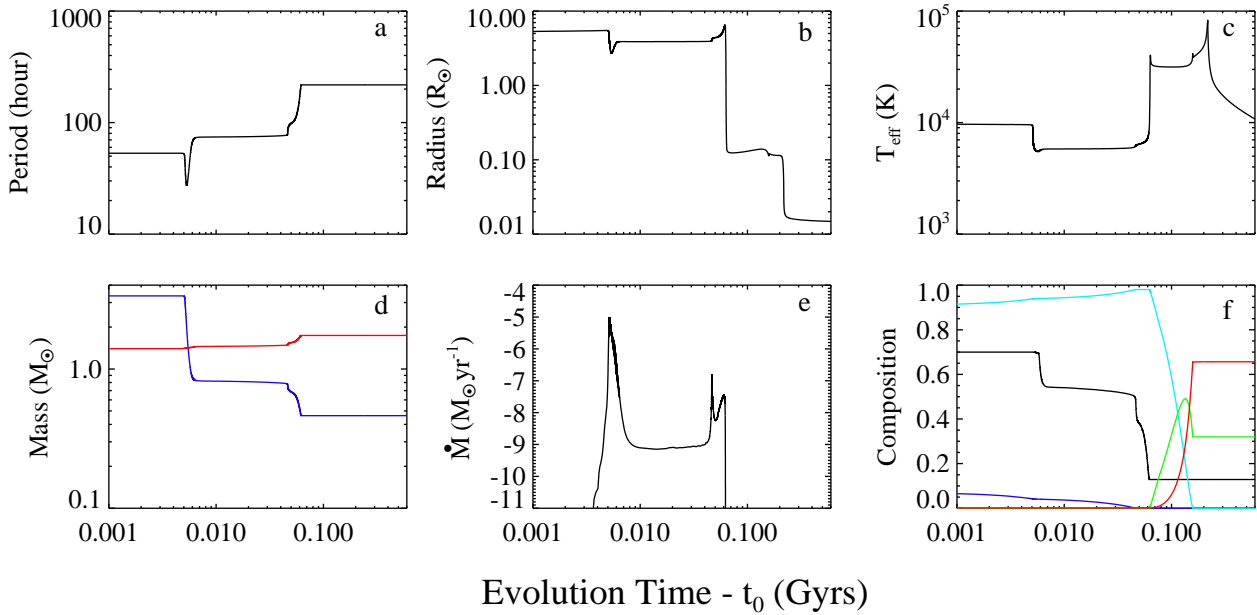} \caption{Detailed evolution vs.\,time of several
parameters associated with the system that best represents the
evolution of PSR J1614-2230 for a NS with initial mass
$1.4\,M_\odot$ (the initial time in the plots, $t_0$, was chosen
as 262 Myr to emphasize the mass-transfer phase). The top panels
show the evolution of $P_{\rm orb}$, $R_2$, and $T_{\rm eff}$ as
functions of time. The bottom panels show the evolution of stellar
mass, $\dot M$, and central composition for the donor star as
functions of time. In the mass panel (d) the blue and red curves
are for the donor star and NS, respectively. In the composition
panel blue, cyan, green, and red are the mass fractions (at the
center of the donor star) of H, He, C12, and O16, respectively.
The black curve is the surface H abundance. }
\label{fig:details_evol}
\end{figure*}

To better understand what the different types of LMXB and IMXB
evolutions contribute to diagram, we show in
Fig.\,\ref{fig:Ransom_like} a set of illustrative individual
tracks, labeled ``CV'', ``UC'', ``G1'', and ``G2''. The CV-like
tracks are analogous to the evolution of cataclysmic variables,
where the donor star is not very evolved at the time when mass
transfer commences, remains H-rich, and gets eaten away by mass
transfer until the orbital period reaches a minimum (at $\sim$70
min.), after which the orbit slowly expands \citep[see, e.g.,][and
references therein]{pacz81,rapp83,howe01}. These systems commence
mass transfer for $P_{\rm orb}$ shorter than the so-called
`bifurcation period' of between $\sim$$20-30$ hrs (depending on
the initial donor mass). The ultracompact systems (``UC'') start
mass transfer in somewhat wider orbits, very near the bifurcation
period, with the donor stars correspondingly more evolved,
typically with H just having been depleted at the stellar center
\citep[see,
e.g.,][]{nels86,tutu87,pyly88,pyly89,fedo89,pods02,nels03}.
Because of the high He content in the core of the donor star,
these systems are able to evolve to much shorter orbital periods
(some as short as 6 min.)  Finally, those systems that commence
mass transfer in even wider orbits (e.g., ``G1'' and ``G2''), lead
to the formation of a well-defined He core and to an increase in
orbital period. For those donor stars which start with $M_2
\lesssim 2.2\,M_\odot$, their degenerate cores prescribe a
mass-radius relation, which in turn dictates an orbital
period-core mass relation, $P_{\rm orb}(M_c)$, which is clearly
delineated by the locus of green terminus points in
Fig.\,\ref{fig:43000} which follows the $P_{\rm orb}(M_c)$
relation \citep[e.g.,][]{rapp95,ergm96,taur99,nels04}.

For donor stars with initial masses $\gtrsim 2.2 \,M_\odot$,
however, this $P_{\rm orb}(M_c)$ relation does not have to be
followed, and their evolution typically terminates considerably
before this $P_{\rm orb}(M_c)$ boundary is reached. This is
compounded by the fact that, due to the high mass ratios involved,
the initial phases of the mass transfer, while dynamically stable,
are governed by the thermal timescale of the donor star (see,
e.g., \citealp{king01,pods00}; PRP02; for further references see
\S \ref{sec:PSRevol}). These rates can therefore become greater
than $10^{-5} M_\odot$ yr$^{-1}$, which is far in excess of the
Eddington limit for a NS\footnote{As high as these rates are, they
typically remain below the rates required for the onset of
`hypercritical accretion' \citep[see, e.g.,][]{houc91} which might
have allowed for the growth of higher mass neutron stars.}.
Therefore, the mass transfer is highly non-conservative, and most
of the mass is ejected from the system. The latter class of
systems, we believe, is responsible for the evolution of both Cyg
X-2 and PSR 1614-2230.

In a forthcoming paper we will concentrate on various other
aspects of these evolutionary paths and describe in greater detail
the various types of LMXBs and recycled pulsars that they lead to.
For the present work we focus on the evolutionary paths leading to
Cyg X-2-like systems and PSR J1614-2230.

\section{Binary Evolution of the PSR J1614-3320 System}
\label{sec:PSR1614}

\subsection{Properties of PSR J1614-3320}

PSR J1614$-$2230 \citep{demo10} is a millisecond radio pulsar in
an $8.7$ day orbit around a (presumably) predominantly
carbon-oxygen (CO) white dwarf ($0.5\pm0.006 \,M_\odot$). The
recent measurement of the Shapiro delay in PSR J1614$-$2230 has
yielded a NS mass of $1.97\pm0.04\,M_\odot$, making it the most
massive pulsar known to date \citep{demo10}. Given the constituent
masses and $P_{\rm orb} \simeq 9$ days, this recycled pulsar
system shows all the signs of having descended from an LMXB very
much like Cyg X-2 ($P_{\rm orb} \simeq 9.8$ days, $M_{\rm NS}
\simeq 1.8\,M_\odot$ and $M_2 \simeq 0.6 \,M_\odot$;
\citealp{casa98,oros99}). The orbital eccentricity of
$\sim$$10^{-6}$ indicates that the evolution involved an extensive
period of mass transfer via a Roche-lobe filling donor star. The
characteristic spin-down age is $\sim$5 Gyr.

\subsection{Evolution Tracks Leading Close to PSR J1614-2230}
\label{sec:PSRevol}

Because of the very interesting and unique system parameters of
PSR J1614-2230, we were motivated to search our grid of tracks for
all those systems which terminate their evolution with $P_{\rm
orb} = 9 \pm 1$ day and $M_2 = 0.5 \pm 0.1\,M_\odot$. In all,
there were 515 such systems; their evolution in the $P_{\rm
orb}-M_2$ plane is shown in Fig.\,\ref{fig:Ransom_like} as light
blue and green tracks. Note that they all start with donor masses
in the range of $3.35 \lesssim M_{\rm 2,i} \lesssim
3.75\,M_\odot$, and $P_{\rm orb,i}$ in the range of $2-4$ days.

Note how the light blue tracks all {\em decrease} in $P_{\rm orb}$
until the donor mass reaches $\sim$$2\,M_\odot$, after which
$P_{\rm orb}$ increases again until mass transfer has ceased when
the donor star no longer has an extended envelope. The initial
decrease in $P_{\rm orb}$ results from the fact that mass is being
transferred from the more massive donor to the less massive NS.
The total duration of the light blue portion of the tracks
corresponds to a characteristic evolution time of only $\sim$ a
Myr. This rapid phase of mass transfer is referred to as ``thermal
timescale'' mass transfer since it takes place on the thermal
timescale of the radiative donor star that is more massive than
the accreting NS (see, e.g., \citealp{pods00}; PRP02; for other
related references see below in \S \ref{sec:PSRevol}). Once the
donor masses reach $\sim$$0.8 \,M_\odot$ the thermal timescale
mass transfer is over, and the evolution slows down to timescales
of tens of Myr - the nuclear evolution timescale of the donors
(see the green and yellow portions of the tracks in
Fig.\,\ref{fig:Ransom_like}).

We have selected one of these 515 tracks, as illustrative, in that
its end product best matches the properties of PSR J1614-2230. We
show details of that particular evolution in
Fig.\,\ref{fig:details_evol}. The various panels show the
evolution of $P_{\rm orb}$, $R_2$, $T_{\rm 2,eff}$, $M_2$, $\dot
M$, and information on the chemical composition of the donor (star
2), as functions of the evolution time. A reference time of 262
Myr has been subtracted from the evolution time to enhance the
short-timescale features around the mass-transfer phases. Note
that $\dot M$ refers to the mass loss rate from the donor star,

The short-lived drop in $P_{\rm orb}$ (panel a) corresponds to the
thermal timescale mass transfer event (the large spike in $\dot M$
in panel e). In panel (f) we see that the central H mass fraction
is only $\sim$6\% by the time mass transfer commences. The
relatively long plateau in $\dot M$ (panel e) at
$\sim$$10^{-9}\,M_\odot$ yr$^{-1}$ which lasts for $\sim$35 Myr,
corresponds to the phase of mass transfer driven by nuclear
evolution during which time the remainder of the H in the core and
much of the envelope is largely consumed. Once a He core has
formed, there is an approximately 4-Myr interval during which
H-shell burning occurs. This causes the radius of the donor star
to expand from $\sim$$3.5\,R_\odot$ to $\sim$$6\,R_\odot$, during
which time the mass transfer rate ranges between
$10^{-8}-10^{-7}\,M_\odot$ yr$^{-1}$. The orbital period increases
from $\sim$3 days to $\sim$9 days, and the donor-star mass drops
to $0.47\,M_\odot$. It is during this second phase of mass
transfer when the accreting NS undergoes its largest growth in
mass. After the H-shell burning phase, the donor star contracts,
Roche lobe contact and the mass-transfer phase is over, and the He
in the core of the donor burns to C and O.   The companion, which
ends up being 90\% C and O (with 10\% of the mass in a He
envelope), contracts to its final degenerate radius of
$0.015\,R_\odot$. This entire evolution is sometimes referred to
as ``case AB'' since the mass transfer occurs near the end of the
main sequence. (For a closely related evolution scenario for Cyg
X-2 see \citealp{king99,pods00}.) and is not the same as the mass
accreted by the NS.

\begin{figure}[t]\epsscale{1.15}
\plotone{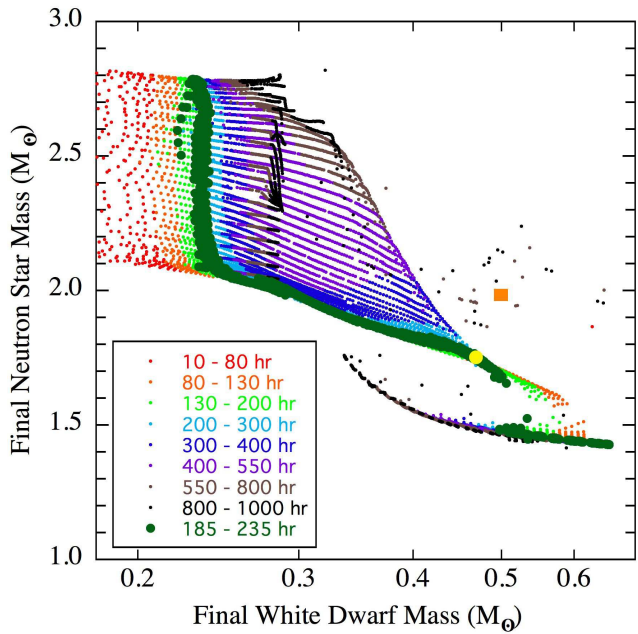} \caption{The final neutron-star mass vs. the
final mass of the white dwarf companion for different ranges of
final $P_{\rm orb}$, as indicated by the color scaling (in hours).
The dark green points are the systems with a final $P_{\rm orb}$
most appropriate to match PSR 1614-2230. The orange square marks
the location of PSR J1614-2230, while the large yellow circle
indicates the system plotted in detail in
Fig.\,\ref{fig:details_evol}. Along the locus of dark green
points, the higher the initial donor mass the {\em lower} the
final neutron star mass. The bottom-most line of systems is due to
case B mass transfer with initial donor masses $\gtrsim
2.2\,M_\odot$. The higher mass neutron stars are very likely
unphysical, and are a consequence of the high mass capture
fraction assumed for sub-Eddington accretion, and not allowing
these stars to collapse to black holes.} \label{fig:NSmass}
\end{figure}

To put the evolution of PSR J1614-2230 into some context, we show
in Fig.\,\ref{fig:NSmass} a plot of the {\em end points} of all
the binary evolutions we ran in the plane of the final
neutron-star mass vs.\,the final donor-star mass (now a white
dwarf) -- with $P_{\rm orb,f}$ restricted to $\gtrsim 10$ hr. In
all, there are some 14,000 system end points represented in this
figure. Even though the points form patterns and lie along pseudo
tracks, it is important to note that they are {\em not} evolution
tracks, but rather {\em end points} of numerous evolutions. These
patterns result from the discrete nature of the grid of starting
values for $P_{\rm orb,i}$ and $M_{\rm 2,i}$. The location of PSR
J1614-2230 is marked with an orange square.

\begin{figure}[t]\epsscale{1.15}
\plotone{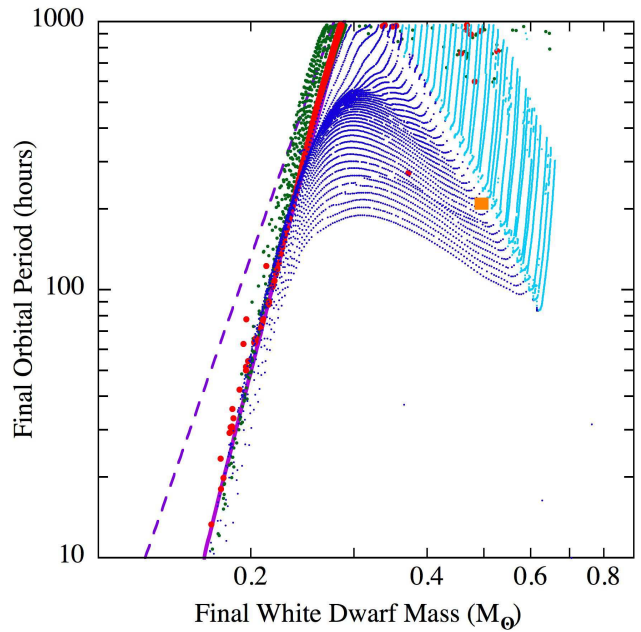} \caption{The final orbital period vs. the final
mass of the white dwarf companion star for $\sim$14,000 systems
whose final orbital period lies in the range of $10-1000$ hr, or
$\sim$1/2 to 40 days. The color coding of the dots is as follows:
cyan and blue are for $M_{\rm 2,i} > 2.2\, M_\odot$ and cases B
and AB, respectively (see text); the green and red represent
$M_{\rm 2,i}$ between $1.4-2.2\, M_\odot$ and $1.0-1.4\,M_\odot$,
respectively. The location of PSR J1614-2230 is marked with an
orange square. The dashed purple line is the short-period, low
mass end of the Rappaport et al.\,1995 $P_{\rm orb}(M_c)$
relation, while the solid purple curve is a fit to the He white
dwarfs (at the left boundary) and also subsumes the Rappaport et
al.\,$P_{\rm orb}(M_c)$ relation for white dwarf masses in the
range of $0.3-1.2\,M_\odot$ which is off this plot. This steeper
slope predicting higher-mass white dwarfs for a given $P_{\rm
orb}$ was earlier alluded to by \citet{taur99}.}
\label{fig:PMplane}
\end{figure}

The end point locations in this $M_{\rm ns,f}-M_{\rm 2,f}$ plane
are color coded according the final value of the orbital period.
The heavy green dots correspond to $P_{\rm orb,f}$ in the range of
$8.7\pm 1$ day, roughly commensurate with $P_{\rm orb}$ of PSR
J1614-2230. The vertical column of heavy green dots originates
from systems with initial donor-star masses $\lesssim
2.2\,M_\odot$ which follow the core-mass radius relation (see
discussion below), and where the mass transfer rates are well
below the Eddington limit. Thus, the NSs in these systems are able
to grow substantially. For higher-mass donors (i.e., $\gtrsim
2.2\,M_\odot$) that commence mass transfer near the end of the
main sequence (case AB), the final NS masses lie in the range of
$\sim$$2.1-1.65\,M_\odot$ with white dwarf masses inversely
correlated with the NS mass, and lying between $\sim$0.26 and
$0.50\,M_\odot$ (the continuing line of heavy green dots). The
line of systems with the most massive white dwarfs (between 0.5
and 0.65 $M_\odot$) arise from case B mass transfer (where the
initial donor masses were $> 2.2 \,M_\odot$), but their companion
NS masses are all $\lesssim 1.5\,M_\odot$.

This `competition' between the NS mass and that of the residual
white dwarf companion is due to the fact that, in general,  the
more the donor star is evolved (hence larger core masses) when
mass transfer commences, the higher is the thermal-timescale
mass-transfer rate, and therefore the lower the mass retention
fraction by the NS (where the Eddington limit is greatly
exceeded). More specifically, the case B mass-transfer systems
have only a thermal-timescale mass-transfer phase, and there is no
sub-Eddington phase during which the NS mass can grow
significantly. Therefore, raising an initial NS mass of
$1.4\,M_\odot$ to $2\,M_\odot$ remains a difficulty for the model
(see also PRP02).

The relation between the final orbital periods and the white dwarf
companion masses of the systems ending with $P_{\rm orb} > 10$
hours is shown in Fig.\,\ref{fig:PMplane}. Again, these are 14,000
{\em end points}, and not evolution tracks. The location of PSR
J1614-2230 is marked. The systems lying along the leftmost
boundary of these end points (marked with a solid line) all
represent He white dwarfs of mass between $\sim$$0.17\,M_\odot$
and $\sim$$0.28\,M_\odot$. These systems form from donor stars of
initial mass $\lesssim 2.2\,M_\odot$, and are well represented by
a nearly unique $P_{\rm orb}(M_c)$ relation \citep[see,
e.g.,][]{rapp95,ergm96,taur99}. The dashed line to the left of
this set of end points is the expression given by \citet{rapp95}
which was already known to systematically somewhat underestimate
the white dwarf mass for systems with short orbital periods below
which the relationship was not designed to work. The solid line is
given by
\begin{equation}
P_{\rm orb} \simeq \frac{4.6 \times 10^6 ~m_c^9}{(1+25 m_c^{3.5} +
29m_c^6)^{3/2}}{\rm ~days} ~~,
\end{equation}
where $m_c$ is the white dwarf mass expressed in solar units. We
devised this expression to fit both the results shown in
Fig.\,\ref{fig:PMplane} for low-mass white dwarfs and to
incorporate the expression of \citet{rapp95} which extends all the
way to white dwarfs of mass $1.2\,M_\odot$.

The systems distinctly to the right of this $P_{\rm orb}(M_c)$
relation originate from donor stars with $M_2 \gtrsim
2.2\,M_\odot$ and do not follow the core mass-radius relation for
giants. Note the ``gap'' in these more massive white dwarfs (which
runs approximately at $\sim$$-45^\circ$) dividing systems that
evolved in so-called case AB (on the left side) from those that
formed from somewhat more evolved donors at the time mass transfer
commences, i.e., case B. These white dwarfs become progressively
more CO-rich in composition toward the higher masses. (For other
related studies see, e.g., \citealp{iben85}; PRP02;
\citealp{taur99,han00,taur00,nels04}.)

Finally, in this regard, we note that PSR J1614-2230 is quite
representative of the systems which started with donors of mass
$\gtrsim 2.2\,M_\odot$ and ended up with white dwarfs of mass in
the range $0.26-0.65\,M_\odot$. The mass transfer rate in case B
systems (toward the right) is so high that the neutron stars do
not have much chance of growing to interestingly high masses. By
contrast, the systems to the left of the `gap' between cases AB
and B have intervals of near- or sub-Eddington rates which allow
the highest mass NSs to be grown (see Fig.\,\ref{fig:NSmass}). In
general, the higher-mass neutron stars tend to be found with the
lower-mass white dwarfs.

\subsection{Neutron Stars with Higher Initial Masses}

As discussed in the previous section, none of our evolution
tracks, starting with neutron stars of canonical mass
$1.4\,M_\odot$, produced the observed $1.97$~M$_{\odot}$ NS with
the requisite  combination of orbital period and white dwarf mass
to match the PSR J1614-2230 system. Neutron stars as massive as
$2\,M_\odot$, and higher, were produced with the correct orbital
period, but not in combination with a massive enough white dwarf.
Similarly, the orbital period and white dwarf mass combination is
easy to reproduce, but not with the correct $P_{\rm orb}$.  We
find that the final values of $M_2$ and $M_{\rm NS}$ are
anticorrelated when $P_{\rm orb}$ is fixed to $8.7 \pm 0.5$ day.
The difficulty with producing high-mass NSs in orbit with massive
white dwarf companions is that the progenitors of these white
dwarfs are initially substantially more massive than the NS,
resulting in very rapid, thermal-timescale mass transfer (at rates
as high as $10^{-5}\,M_\odot$ yr$^{-1}$). Unless this is followed
by a substantial interval of sub- or near-Eddington accretion
rate, the neutron star will likely be prevented from accreting a
significant fraction of the donor star's mass.

\begin{figure}[t]\epsscale{1.17}
\plotone{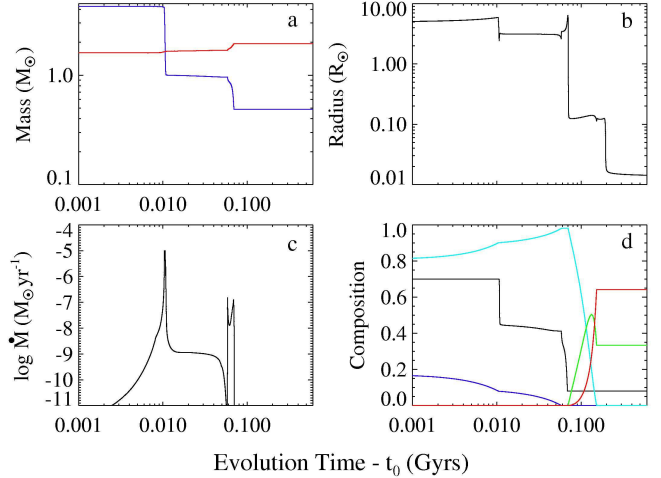} \caption{Detailed evolution vs.\,time of
several parameters associated with the system that matches the
parameters of PSR J1614-2230 for a NS with initial mass
$1.6\,M_\odot$ and donor star of initial mass of $4.25\,M_\odot$.
As in Fig.\,\ref{fig:details_evol} $t_0$ is an initial time that
is subtracted off so as to emphasize the mass-transfer phase. The
top panels show the evolution of the stellar masses, and the
radius of the donor star, while the bottom panels show the
evolution of $\dot M$, and the central composition of the donor
star.  The color coding in the mass curves and chemical
composition are the same as in Fig.\,\ref{fig:details_evol}. The
final masses of the neutron star and white dwarf in this sequence
were 1.95 $M_\odot$ and 0.49 $M_\odot$, respectively, and the
final orbital period (not shown) was 206 hours (8.4 days).}
\label{fig:success}
\end{figure}

One obvious solution to this problem of producing more massive NSs
is to start with higher natal mass neutron stars.   To this end,
we have run a series of smaller subgrids of 700 additional
starting models with initially higher-mass neutron stars.  We find
that the minimum required initial mass neutron star for which we
could reach a final NS mass of $1.97 \,M_\odot$ turned out to be
$1.6 \, M_\odot$.  The starting donor mass for these more
`successful' models, however, increased to $\sim$$4.25 \pm 0.10
\,M_\odot$ (compared to  $\sim$$3.8 \,M_\odot$), the initial
orbital periods remained near $\sim$$49 \pm 2$ hours, and the
final companion CO white dwarf mass was $\sim$$0.49 \,M_\odot$,
somewhat closer to the observed value.   An illustrative evolution
sequence that produces a $\sim$$2 \,M_\odot$ NS, and otherwise
closely resembles PSR J1614-2230, is shown in
Fig.\,\ref{fig:success}.

Finally, in regard to setting a limit on the mass of the natal
neutron star, we note an important caveat.  Throughout our
calculations we adopted a very simple prescription for the
Eddington-limit rate of accretion onto the neutron star of $\dot
M_{\rm Edd} = 3 \times 10^{-8} \, M_\odot$ yr$^{-1}$, independent
of the neutron star mass or radius, or the chemical composition of
the accreted material.  For most of the evolutionary phases we
covered in our 42,000 tracks, the accretion rate is either below
$10^{-8} \, M_\odot$ yr$^{-1}$, in which case the value we choose
for $\dot M_{\rm Edd}$ is unimportant, or it so high, during the
thermal timescale mass-transfer phases, that the choice of $\dot
M_{\rm Edd}$ also does not affect the very small fraction of mass
that can be retained by the neutron star.  However, for the
production of systems that resemble PSR J1614-2230, with its
massive neutron star and other specific properties, the largest
growth of the neutron star occurs during the second phase of mass
transfer in case AB evolution where $\dot M$ is in the critical
range of $10^{-8} - 10^{-7} \, M_\odot$ yr$^{-1}$, and where the
value of $\dot M_{\rm Edd}$ very much affects how much the neutron
star can grow.  Adjusting the value of $\dot M_{\rm Edd}$ to be
``more correct'' is complicated by the following issues: (i) the
unknown radius of the NS (by a factor of $\sim$$50\%$), (ii)
general relativistic corrections (of $\sim$20\%), (iii) unknown
allowed factors by which the Eddington limit may be violated in NS
accretion (of $\sim$ factors of 2), and (iv) changes in chemical
composition of the accreted material -- which is especially
important as the material becomes He-rich later in the evolution.
Also, the amount of mass gained by the NS is only some $\sim$80\%
of what was transferred through the accretion disk due to the loss
of rest mass that is radiated away.  Taking all these factors into
account, we have run a range of tests and have concluded that such
effects and their uncertainties, lead to an uncertainty in the
minimum mass for a neutron star to reach 1.97 $M_\odot$ of
$\sim$$\pm 0.1\,M_\odot$.  Therefore, the range of required
minimum neutron star masses could conceivably be expanded to $1.6
\pm 0.1$ $M_\odot$.

\section{Summary and Conclusions}
\label{sec:Discuss}

We have used {\tt MESA} to compute a large grid of 42,000 binary
evolution models for LMXBs and IMXBs. We showed the broad sweep of
possible evolutions, from systems which attain orbital periods as
short as 6 minutes to those which grow to long-period binaries
with giant donor stars. We leave a detailed discussion of the bulk
of these results for a future paper. Here we have focused on what
we can learn about the evolutionary paths to the newly discovered
binary pulsar, PSR J1614-2230. We have selected a subset of the
evolution models (515 in total) which lead to systems like PSR
J1614-2230 and its evolutionary cousin, Cyg X-2, to examine in
more detail. In particular, we show how an orbital period of
8.7-days can be easily understood, as can the $0.5\,M_\odot$
companion star. We show that for the proposed scenario, the
degenerate companion star is mostly (i.e., 90\%) C and O (with a
surrounding shell of He which comprises 10\% of the white-dwarf
mass), but there is also a thin outer envelope that is composed of
$\sim$15\% H.

We found, however, that starting with neutron stars of canonical
mass $1.4\,M_\odot$, we were not able to produce the observed
$1.97$~M$_{\odot}$ NS with the requisite  combination of orbital
period {\it and} white dwarf mass to match the PSR J1614-2230
system.  We were able to evolve neutron stars as massive as
$2\,M_\odot$, with the correct orbital period, but not in
combination with a massive enough white dwarf.  Also, the orbital
period and white dwarf mass combination was easy to generate, but
not with the correct $P_{\rm orb}$.  As discussed above, the
difficulty with evolving high-mass NSs in orbit with massive white
dwarf companions is that the progenitors of these white dwarfs are
initially substantially more massive than the NS, resulting in
very rapid, thermal-timescale mass transfer (greatly in excess of
the Eddington limit), and the neutron star is thereby prevented
from accreting a significant fraction of the donor star's mass.

At this time, we are therefore tentatively led to conclude that
the initial mass of the NS would have to have been higher than the
canonical value of $1.4\,M_\odot$ in order to declare that the
binary evolution of this system is fully understood. By running
some 700 supplementary models with initially higher-mass NSs, we
found that to successfully produce a system like PSR J1614-2230
requires a minimum initial neutron star mass of at least $1.6 \pm
0.1 \, M_\odot$, as well as initial donor masses and $P_{\rm orb}$
of $\sim$$4.25 \pm 0.10 \,M_\odot$ and $\sim$$49 \pm 2$ hrs,
respectively.

For completeness in regard to producing high neutron-star masses,
we point out that a number of the dynamically stable case B
evolution tracks that we have generated yield mass transfer rates
in excess of several times $10^{-4}\,M_\odot$ yr$^{-1}$ that last
for intervals of several thousand years. If `hypercritical
accretion', where the gravitational energy is carried off in
neutrinos and the Eddington limit is thereby circumvented, is able
to occur during these intervals \citep[see,
e.g.,][]{houc91,brow00}, then perhaps there is a chance for the
neutron stars to grow during this phase of the evolution. However,
it is not clear to us that the hypercritical accretion scenario,
which was developed for spherical accretion onto a neutron star,
would be applicable to accretion via a disk \citep[but,
see][]{more08}.

A more general conclusion from this L/IMXB study is that there is
a subclass of systems which start with intermediate mass donor
stars (of $\gtrsim 2.2 \, M_\odot$), with $P_{\rm orb}$ well above
the `bifurcation period' (in the range of $2-4$ days), which leads
to systems like Cyg X-2 and PSR J1614-2230. Such systems terminate
their mass transfer before the donor star develops a degenerate
core, and they end up with $P_{\rm orb}$ well below the $P_{\rm
orb}(M_c)$ relation. (See also \citealp{iben85}; PRP02;
\citealp{han00,taur00}).

Another important general lesson that we can take from this broad
look at L/IMXB evolution is that {\it intermediate}-mass donor
stars  can evolve to virtually all the known types of LMXB systems
that exist at the current epoch. These include CV-like evolution
paths, ultracompact X-ray binaries, and systems with giant donor
stars. This is an important finding because it is significantly
easier for intermediate-mass stars to successfully eject the
envelope of their massive companion progenitors of the NSs, and
then remain bound during the ensuing supernova explosion.

Finally, we have provided a perspective on where in the $P_{\rm
orb}$ and white dwarf mass plane we can expect to find the He
white dwarfs that follow the $P_{\rm orb}(M_c)$ relation, as well
as where the systems with the most massive neutron stars should be
found. Depending on the upper mass limit to a neutron star,
$M_{\rm ns, max}$, Fig.\,\ref{fig:NSmass} shows that there should
be a fair number of black holes with mass between $M_{\rm ns,
max}$ and $\sim$$2.8\,M_\odot$ in binaries with He or He/CO white
dwarfs that range between $\sim$0.2 and 0.4 $M_\odot$ and with
orbital periods in the range of $1-40$ days. One way to detect
such important relics of stellar evolution is to search for white
dwarfs with interestingly high orbital velocities (i.e., $v \simeq
300 \,(P_{\rm orb}/{\rm days})$ km s$^{-1}$) and unseen
companions. An important caveat to this is that if such low-mass
black holes form in significant numbers in LMXBs that are still
undergoing mass transfer, they would be directly detected by their
accretion.  However, there is no evidence for LMXBs with low-mass
black holes (see, e.g., \citealp{farr10}; \citealp{ozel10}).

\acknowledgements We thank Deepto Chakrabarty and Lars Bildsten
for helpful discussions, and an anonymous referee for very
constructive suggestions. We acknowledge the early participation
in this work by Nikku Madhusudhan and Josiah Schwab. J.L. grateful
acknowledges support from NASA {\em Chandra} grant G09-0054X; B.P.
thanks the NSF for support through grants PHY 05-51164 and AST
07-07633 (BP); and SR was supported in part by NASA {\em Chandra}
grant TM8-9002X. We acknowledge the R\'eseau qu\'eb\'ecois de
calcul de haute performance (RQCHP) for providing the
computational facilities and we thank the Natural Sciences and
Engineering Research Council of Canada and the Canada Research
Chairs progam for financial support (LN).

\bibliographystyle{apj}

\end{document}